\documentclass[aps,floatfix,showpacs,showkeys,amsmath]{revtex4}
\usepackage{color,graphicx}
\usepackage{dcolumn}
\usepackage{bm}
\usepackage{epsfig}
\usepackage{supertabular}
\usepackage[bookmarksopen]{hyperref}
\def\be {\begin{equation}}
\def\ee {\end{equation}}
\def\nn {\nonumber}
\def\bea {\begin{eqnarray}}
\def\eea {\end{eqnarray}}

\begin{document}

 \title{Wake potential in collisional anisotropic quark-gluon plasma }
\bigskip
\bigskip
\author{ Mahatsab Mandal}
\email{mahatsab.mandal@saha.ac.in}
\author{ Pradip Roy}
\email{pradipk.roy@saha.ac.in}
\affiliation{Saha Institute of Nuclear Physics, 1/AF Bidhannagar
Kolkata - 700064, India}

\begin{abstract}
Within the framework of Boltzmann transport equation with a 
Bhatnagar-Gross-Krook (BGK) collisional kernel,
we study the wake potential induced by fast partons 
traveling through the high-temperature QCD plasma 
which is anisotropic in momentum-space. 
We calculate the dielectric response function
of a collisional anisotropic quark-gluon plasma (AQGP) for 
small $\xi$ (anisotropic parameter) limit. Using this, the wake 
potential for various combinations of the anisotropy parameter ($\xi$)
and the collision rate ($\nu$) is evaluated both for parallel and perpendicular
directions of motion of the fast parton. It is seen that the inclusion of the
collision modifies the wake potential and the amount as well as the nature
of the potential depends on the combinations of $\xi$ and $\nu$.
\end{abstract}
\maketitle

\section{Introduction}
The aim of ultrarelativistic heavy-ion collision experiments at 
BNL RHIC and at CERN LHC is to understand the properties 
of very hot and dense partonic matter expected to be formed in
these collisions. One of the objectives 
of such experiments is the identification and investigation
of a phase transition from hadronic matter to what is known as 
the quark-gluon plasma (QGP). 
The hard jets created in hard parton scattering in the 
initial stages of heavy-ion collision will pass through
the hot and dense matter and lose energy by collisional
and radiative processes. This phenomenon is known as jet 
quenching where the high $p_T$ hadrons production has been found 
to be strongly suppressed~\cite{jetquen1}. Other very 
important experimental evidence for the jet energy loss is 
the dihadron azimuthal correlation. The jet induced hadron
pair distribution at RHIC shows a double peak structure in 
the away side~\cite{prl95,prl97} for the intermediate 
$p_T$ particle. However, it is unclear how the 
parton-medium interaction affects the distribution.
The possible explanation of this  observation is that  the 
coupling of jets to a strongly interacting medium may modify
the angular distribution~\cite{prl90,prc70,jconf,jpg_34,npa750,
prl96,prc73,plb618,npa774}. In this scenario the partonic jets  
traveling through the QGP leads to formation of Mach cones
~\cite{jconf,jpg_34,npa750}, Cerenkov radiation~\cite{prl96, prc73}
and wakes~\cite{plb618,prd74,jpg34,npa856,prd86}, which may
be observables as collective excitation of the medium.

Apart from the jet quenching many other signals of this nascent
phase of matter have been proposed. These include electromagnetic probes
(photon and dilepton)~\cite{annals},
$J/\psi$ suppression~\cite{matsui}, collective 
flow~\cite{Pasi,Hirano,Tannenbaum} and so on.
However, the main difficulty in studying these probes lies in the
determination of the initial conditions, such as, the 
isotropization/thermalization time vis-a-vis the initial
temperature.
In this sense, many  properties of the QGP are poorly 
understood. The most pertinent question is whether the 
matter produced in relativistic heavy-ion collisions is in 
thermal equilibrium or not. The collective flow patterns 
observed at the RHIC provide strong evidence for rapid thermalization 
at less than 1 fm/c after the collision~\cite{arxive0512051}. 
On the other hand, using second-order transport coefficients 
consistent~\cite{prc_78} with conformal symmetry it has been found 
that the thermalization time has sizable uncertainties due to poor 
knowledge of the proper initial conditions. Plasma instabilities 
have been suggested to a play a major role in the isotropization
process~\cite{6ofprc78}. Shortly after the collision, the rapid
expansion of the matter along the beam direction causes faster
cooling in the longitudinal direction than  the transverse 
direction, leading to 
$\langle{p_L}^2\rangle << \langle{p_T}^2\rangle$~\cite{plb502}
i.e., the phase space distributions of plasma particles become
anisotropic in momentum space. 
At some later time, the
system returns to an isotropic state due to the effect of the 
parton interactions which overcomes the plasma expansion rate. 
Therefore, it has been suggested to look for some observables which
are sensitive to the early time after the collision. The effect of
pre-equilibrium momentum anisotropy on various observables has been
studied extensively in the last few years. The collective modes in an
AQGP has been investigated in Refs.~\cite{prd68,prd70}. The energy 
loss of parton in an AQGP has been studied in 
Refs.~\cite{prd71,prc83,prc84}. The effect of anisotropy on the photon and 
dilepton yield have been investigated rigorously in 
Refs.~\cite{prd76,prl100,prc_78_mm,prc78,prc79}. 

When a high energy jet propagates through plasma, apart from the 
energy loss, it produces a wake in the plasma. The current and 
charge density wakes induced by a jet propagating in an isotropic
and homogeneous QGP have been investigated by Ruppert and Muller
~\cite{plb618}. In the high temperature regime, the result shows 
the wake in both the induced charge and current density due to 
the screening color charge cloud, but no Mach cones appear. 
On the other hand, in the quantum liquid scenario, if the jet 
travels supersonically, the wake exhibits an oscillatory behavior 
and Mach cones can appear. However, using the HTL approximation, 
the wake in the induced charge density shows an oscillatory 
nature in the backward direction at large parton 
velocity~\cite{prd74} in case of isotropic plasma. The color 
response wake in viscous media has been investigated in 
Ref.~\cite{npa856}. In a previous paper~\cite{prd86} we studied 
the wake induced by the jet propagating through the 
anisotropic plasma. When the parton moves along the anisotropy 
direction with velocity less than the average plasmon speed 
($v_p$), the anisotropy effect causes a small oscillation 
of the induced charge density. Anisotropy amplifies the 
oscillatory behavior of the wake potential in the backward 
space when parton moves along the beam direction with velocity 
$v > v_p$. However, we do not find any oscillatory nature of 
the wake potential, when the parton moves in the transverse plane
with respect to the anisotropy axis. 
In all the above works the effect of collision has been neglected.
The inclusion of collision in the Boltzmann equation and its
subsequent effect on the wake potential has been investigated
in Ref.~\cite{jpg34} assuming isotropic plasma. It is found that
in case of collisional isotropic plasma, the wake behavior is less
pronounced than in isotropic plasma~\cite{jpg34}.
Now we investigate the wake behavior of the plasma by taking 
into account collisions in the AQGP. The effect of the collision
can easily be taken into account by approximating the collision
term in the Boltzmann equation with the BGK description~\cite{pr94}. 
In this approach, the dielectric function has been calculated in 
isotropic plasma~\cite{canj82} and the collective modes of 
anisotropic collisional plasma~\cite{prd73} have also been 
investigated when the wave vector is parallel to the 
anisotropy direction. By using the BGK collisional kernel, we 
calculate the dielectric response function of a collisional 
AQGP for small anisotropy limit. Based on this dielectric 
function, we will investigate the wake potential induced by 
a fast parton traveling through the collisional anisotropic medium. 

The organization of the paper is as follows. In Sec. II, we derive 
the dielectric response function for collisional anisotropic QGP.
In Sec. III we use the dielectric function to determine the wake 
potential in collisional AQGP and discuss the numerical results.
Finally, we conclude in Sec. IV. 

\section{Self-Energy in collisional anisotropic plasma}
To obtain the gluon polarization tensor of a collisional QGP,
we start from the Boltzmann transport equation:
\be
V.\partial_X\delta f^i_a(p,X)+g\theta^i V_{\mu}F^{\mu\nu}_a(X)
\partial_{\nu}^{(p)}f^i({\bf p}) = C^i_a(p,X),
\ee
where $V^{\mu} = (1, {\bf p}/|{\bf p}|)$, $\theta^q = \theta^g = 1$ and 
$\theta^{\bar q} = -1$. 
$F^{\mu\nu} = \partial^{\mu}A^{\nu} - \partial^{\nu}A^{\mu} -i g [A^{\mu}, A^{\nu}]$
is the gluon field strength tensor with gauge field $A^{\mu} = A^{\mu}_aT^a$ or $A^{\mu} = A^{\mu}_a\tau^a$,
where $\tau^a$, $T^a$ with $a = 1, ...N_c^2-1$ are the $SU(N_c)$ group 
generators in the fundamental and adjoint representations with 
${\rm Tr}[\tau^a,\tau^b] = \frac{1}{2}\delta^{ab}$, ${\rm Tr}[T^a,T^b] = N_c\delta^{ab}$
and $g$ is the coupling constant. 
The scalar functions $\delta f^i_a(p,X)$ are found by the projections:~\cite{prd73}
\bea
\delta f^{q/{\bar q}}_a(p,X) &=& 2 {\rm Tr}[\tau_a n^{q/{\bar q}}(p,X)],\nn\\
\delta f^g_a(p,X) &=& \frac{1}{N_c}{\rm Tr}[T_a n^g(p,X)],
\eea
with the colorless  background fields $n^{q/{\bar q}}({\bf p})=f^{q/{\bar q}}({\bf p})I$ and 
$n^g({\bf p}) = f^g({\bf p}){\cal I}$. $I$ and ${\cal I}$ are unit
matrices in the fundamental and adjoint representation, respectively.
The scalar functions $f^i_a({\bf p})$
are also found by projections:
\bea
f^{q/{\bar q}}(\bf p) &=& \frac{1}{N_c}{\rm Tr}[n^{q/{\bar q}}(p,X)],\nn\\
f^g(\bf p) &=& \frac{1}{N_c^2 -1}{\rm Tr}[n^g(p,X)].
\eea
In this paper, we will concentrate on the soft scale, i.e., $k\sim g T <<T$.
The gauge field fluctuation is $A\sim \sqrt{g}T$ and the derivatives 
are of the order $\partial_X\sim g T$. We can neglect the higher order
coupling constant and correspondingly $D_X\rightarrow\partial_X$ and 
$F^{\mu\nu} = \partial^{\mu}A^{\nu} - \partial^{\nu}A^{\mu}$. 
The BGK collisional term $C^i_a$~{\cite{pr94}} can be represented as 
\be
C^i_a(p,X) = -\nu\big[f^i_a(p,X)-\frac{N^i_a(X)}{N^i_{eq}}f^i_{eq}(|{\bf p}|)\big]
\ee
with $f^i_a(p,X) = f^i({\bf p}) + \delta f^i_a(p,X)$. This BGK collision
term corresponds to an improvement of the relaxation time approximation
for the collision term of the Boltzmann equation~\cite{canj82}. The collision
rate $\nu$ is independent of velocity and particle species. We define the particle
number as,
\be
N^i_a(X) = \int_{{\bf p}}f^i_a(p,X),~~~~~N^i_{eq} = \int_{{\bf p}}f^i_{eq}(|{\bf p}|) 
=\int_{{\bf p}}f^i({\bf p}), 
\ee
with $\int_{{\bf p}} = \int \frac{d^3p}{(2\pi)^3}$.
Thus the  transport equation reads as:
\be
V.\partial_X\delta f^i_a(p,X)+g\theta^i V_{\mu}F^{\mu\nu}_a(X)
\partial_{\nu}^{(p)}f^i({\bf p}) = -\nu\Bigg[f^i({\bf p}) + 
\delta f^i_a(p,X) -\Big(1 + \frac{\int_{{\bf p}}f^i_a(p,X)}
{\int_{{\bf p}}f^i({\bf p})}\Big)f^i_{eq}(|{\bf p}|)\Bigg]
\ee
After Fourier transformation of $\delta f^i(p,K)$ and $F^{\mu\nu}(K)$,
the above equation can be written as:
\bea
\delta f^i(p,K) &=& \frac{-ig\theta^i V_{\mu}F^{\mu\nu}(K)\partial_{\nu}^{(p)}f^i({\bf p})
+i\nu(f^i_{eq}({\bf p})-f^i({\bf p}))+i\nu f^i_{eq}({\bf p})(\int_{{\bf p}^{\prime}}
\delta f^i(p^{\prime},K))/N_{eq}}{\omega-{\bf k}.{\bf v}+i\nu}\nn\\
&=&\delta f_0^i(p,K)+i\nu D^{-1}(K,{\bf v},\nu)\frac{f^i_{eq}({\bf p})}{N_{eq}}
\int_{{\bf p}^{\prime}}\delta f_0^i(p^{\prime},K)\nn\\
&+& i\nu D^{-1}(K,{\bf v},\nu)\frac{f^i_{eq}({\bf p})}{N_{eq}}\frac{i\nu}{N_{eq}}
\int_{{\bf p}^{\prime}}f^i_{eq}({\bf p}^{\prime})D^{-1}(K,{\bf v}^{\prime},\nu)
\int_{{\bf p}^{\prime\prime}}\delta f^i_0(p^{\prime\prime},K) + ...\nn\\
&=&\delta f_0^i(p,K)+i\nu D^{-1}(K,{\bf v},\nu)\frac{f^i_{eq}({\bf p})}{N_{eq}}
\zeta(K)\frac{1}{1-\rho},
\eea
where 
\be
\delta f_0^i(p,K) = \big[-ig\theta^i V_{\mu}F^{\mu\nu}(K)\partial_{\nu}^{(p)}f^i({\bf p})
+i\nu(f^i_{eq}({\bf p})-f^i({\bf p}))\big]D^{-1}(K,{\bf v},\nu)
\ee
with $D^{-1}(K,{\bf v},\nu) = \omega - {\bf k}.{\bf v}+i\nu$, 
$\zeta(K) = \int_{{\bf p}} \delta f^i_0(p,K)$ and 
$\rho(K,\nu) = \frac{i\nu}{N_{eq}}\int_{{\bf p}}f^i_{eq}({\bf p})D^{-1}(K,{\bf v},\nu)$.

The induced current for each particle species $i$ is
\bea
J^i_{ind~a}(K) &=& g\int_{{\bf p}}V^{\mu}\delta f^i_a(p,K)\nn\\
&=& g^2\int_{{\bf p}}V^{\mu}\partial^{\beta}_{(p)}f^i({\bf p})
{\cal M}_{\gamma\beta}(K,V)D^{-1}(K,{\bf v},\nu)A^{\nu}_a(K) + g\nu{\cal S}^i(K,\nu)\nn\\
&+& g^2\frac{i\nu}{N^i_{eq}}\int_{{\bf p}}V^{\mu}f^i_{eq}(|{\bf p}|)D^{-1}(K,{\bf v},\nu)
\Big[\int_{{\bf p}^{\prime}}\partial^{\beta}_{(p)}f^i({\bf p}^{\prime})
{\cal M}_{\gamma\beta}(K,V^{\prime})D^{-1}(K,{\bf v}^{\prime},\nu)A^{\nu}_a(K)\nn\\
&+& g\nu{\cal S}^i(K,\nu)\Big]{\cal W}_i^{-1}(K,\nu),
\eea
with ${\cal M}_{\gamma\beta}(K,V) = g_{\gamma\beta}(\omega-{\bf k}.{\bf v})-V_{\gamma}K_{\beta}$,
 ${\cal S}^i(K,\nu) = \theta_i\int_{{\bf p}}V^{\mu}[f^i({\bf p})-
f^i_{eq}(|{\bf p}|)]D^{-1}(K,{\bf v},\nu)$ and\\
${\cal W}_i(K,\nu) = 1-\frac{i\nu}{N^i_{eq}}
\int_{{\bf p}}f^i_{eq}(|{\bf p}|)D^{-1}(K,{\bf v},\nu)$.
The total induced current is given by 
\be
J^{\mu}_{ind~a}(K) = 2N_c J^{g~\mu}_{ind~a}(K) + 
N_f[J^{q~\mu}_{ind~a}(K)+J^{{\bar q}~\mu}_{ind~a}(K)]
\ee
Assuming the  same distribution functions for the quarks and anti-quarks, the total
induced current can be written as 
\bea 
J^{\mu}_{ind~a}(K) &=& g^2\int_{{\bf p}}V^{\mu}\partial^{\beta}_{(p)}f({\bf p})
{\cal M}_{\gamma\beta}(K,V)D^{-1}(K,{\bf v},\nu)A^{\gamma}_a(K) + 2 N_cg \nu {\cal S}^g(K,\nu)\nn\\
&+& g^2(i\nu)\int\frac{d\Omega}{4\pi}V^{\mu}D^{-1}(K,{\bf v},\nu)
\int_{{\bf p}^{\prime}}\partial^{\beta}_{(p)}f({\bf p}^{\prime})
{\cal M}_{\gamma\beta}(K,V^{\prime})D^{-1}(K,{\bf v}^{\prime},\nu){\cal W}^{-1}(K,\nu)A^{\nu}_a(K)\nn\\
&+& 2iN_cg^2\nu^2\int\frac{d\Omega}{4\pi}V^{\mu}D^{-1}(K,{\bf v},\nu)
S^g(K,\nu){\cal W}^{-1}(K,\nu).\label{currentdensity}
\eea
where ${\cal W}(K,\nu) = 1-i\nu\int\frac{d\Omega}{4\pi}D^{-1}(K,{\bf v},\nu)$.
From this expression of the total induced current the self energy 
can be obtained via
\be
\Pi^{\mu\nu}_{ab}(K) = \frac{\delta J^{\mu}_{ind~a}(K)}{\delta A^b_{\nu}(K)}
\ee
which gives 
\bea
\Pi^{\mu\nu}_{ab}(K) &=& \delta_{ab}g^2\int_{{\bf p}} V^{\mu}\partial_{\beta}^{(p)}f({\bf p})
{\cal M}^{\gamma\beta}(K,V)D^{-1}(K,{\bf v},\nu)+\delta_{ab}ig^2\nu
\int\frac{d\Omega}{4\pi}V^{\mu}D^{-1}(K,{\bf v},\nu)\nn\\
&\times&\int_{{\bf p}^{\prime}}\partial_{\beta}^{(p)}f({\bf p}^{\prime})
{\cal M}^{\gamma\beta}(K,V^{\prime})D^{-1}(K,{\bf v}^{\prime},\nu){\cal W}^{-1}(K,\nu),
\label{polarization}
\eea
which is a symmetric tensor, i.e., $\Pi^{\mu\nu}(K) = \Pi^{\nu\mu}(K)$ and
transverse, i.e., $K_{\mu}\Pi^{\mu\nu}(K) = 0$.
Since $\Pi^{\mu\nu}$ is gauge invariant one can evaluate
it in any gauge. In the following we shall evaluate it in the 
temporal axial gauge where $A_0 = 0$.

The anisotropic phase-space distribution function can 
be obtained from any isotropic distribution function by 
rescaling only in one direction in momentum space by 
changing the argument ~\cite{prd68,prd62}
\be
f({\bf p}) = f_{\xi}({\bf p}) = f_{iso}({\bf p}^2+\xi({\bf p.\hat{n}})^2),
\ee
where $\xi$ is an adjustable parameter which represents
the strength of the anisotropy and the direction of 
anisotropy is determined by ${\bf \hat n}$ (assumed to be in the
beam direction). $f_{iso}$ is an 
arbitrary isotropic distribution function. Using this 
distribution function, the gluon polarization tensor of a 
collisional QGP within the BGK approach, is given by~\cite{prd73} 
\bea
\Pi^{ij}(K) &=& m_D^2\int \frac{d\Omega}{4\pi} v^i\frac{v^l+\xi({\bf v.\hat n})n^l}{1+\xi({\bf v.\hat n})^2}
\big[\delta^{jl}(\omega-{\bf k.v}) + v^jk^l\big]D^{-1}(K,{\bf v},\nu)\nn\\
&+& i\nu m_D^2\int \frac{d\Omega^{\prime}}{4\pi}(v^{\prime })^iD^{-1}(K,{\bf v}^{\prime},\nu)
\int\frac{d\Omega}{4\pi}\frac{v^l+\xi({\bf v.\hat n})n^l}{(1+\xi({\bf v.\hat n})^2)^2}
\big[\delta^{jl}(\omega-{\bf k.v}) + v^jk^l\big]D^{-1}(K,{\bf v},\nu)W^{-1}(K,\nu)
\eea 
where $m_D$ is the isotropic Debye mass, represented by
\be
m_D^2 = -\frac{g^2}{2\pi^2}\int^{\infty}_0 dp p^2\frac{df_{iso}(p^2)}{dp}.
\ee 
In the limit $\nu\,\rightarrow\,0$, the polarization
tensor reduces to the  self-energy in the anisotropic medium.
Using the proper tensor basis~\cite{prd68}, one can 
decompose the self-energy into four structure functions as 
\be
\Pi^{ij} = \alpha A^{ij} + \beta B^{ij} + \gamma C^{ij} + \delta D^{ij} 
\ee
 where 
\bea
A^{ij} &=& \delta^{ij} - \frac{k_ik_j}{k^2}, ~~~~~B^{ij} = \frac{k_ik_j}{k^2},\nn\\
C^{ij} &=& {\tilde n}^i{\tilde n}^j, ~~~~~~~~~~~~D^{ij} = k^i{\tilde n}^j + k{\tilde n}^ik^j
\eea
with ${\tilde n}^i = A^{ij}n_j$ which obeys 
${\tilde n}.k = 0$ and $n^2 = 1$. 
The structure functions
$\alpha$, $\beta$, $\gamma$ and $\delta$ are determined 
by the following contractions:
\bea
k_i\Pi^{ij}k_j&=&{\bf k}^2\beta,~~~~~~~~~
\tilde{n}_i\Pi^{ij}k_j=\tilde{n}^2{\bf k}^2\delta,\nonumber\\
\tilde{n}_i\Pi^{ij}\tilde{n}_j&=&\tilde{n}^2(\alpha+\gamma),~~~~~
{\rm Tr}\Pi^{ij}=2\alpha+\beta+\gamma.
\eea 
The structure functions depend on $\omega$, ${\bf k}$, 
anisotropy parameter($\xi$), collisional rate($\nu$) 
and the angle ($\eta$) between the anisotropy vector and the 
momentum. Now the dielectric tensor and the 
self-energy are related by the following relation 
(see Appendix A for derivation):
\be 
\epsilon^{ij}=\delta^{ij}-\frac{\Pi^{ij}}{\omega^2}\label{di-self}
\ee
Using the relation 
$\epsilon({\bf k},\omega)=\frac{k_i\epsilon^{ij}({\bf k},\omega)k_j}{k^2}$~\cite{ichimaru}, 
we can calculate the dielectric function ($\epsilon({\bf k},\omega)$) analytically 
for small $\xi$ limit. To linear order in $\xi$ we have from Eqs.(2) and (7) 
\bea
\epsilon({\bf k},\omega) &=& 1- \frac{m_D^2}{k}
\frac{1}{k-\frac{i\nu}{2}\ln(\frac{z+1}{z-1})} \Bigg[\Sigma+\xi\Big[\frac{1}{6}(1+3\cos 2\eta)+
\Sigma\Big(\cos 2\eta-\frac{z^2}{2}(1+3\cos 2\eta)\Big)\Big]\Bigg]
\eea 
where $z = \frac{\omega+i\nu}{k}$ and 
\be
\Sigma = -1 + \frac{z^2}{2}\ln(\frac{z+1}{z-1}). 
\ee
Note that since $\epsilon$ has been extracted from the polarization
tensor that is gauge invariant, it is indeed gauge invariant also.
The real and imaginary parts of the dielectric function read as
\bea
{\rm Re}\,\epsilon(\omega,{\bf k}) &=& 1 + \frac{m_D^2}{4k^2 G}\Big[4k^2 + \nu^2(\ln^2R+\Theta^2)
-2k(\omega\ln R+2\nu\Theta)\Big]\nn\\
&+& \frac{m_D^2\xi}{24k^4G}\Bigg[-4k^4 + 2k^3\nu\Theta + 12k^2(\nu^2-\omega^2) + 
6k^2\cos 2\eta\Big(2k^2-6\omega^2-k(2\omega\ln R + 3\nu\Theta)\nn\\
&+& \nu^2(6+\ln^2 R+\Theta^2)\Big) +
3(1+3\cos 2\eta)\Big(\nu^2(\ln^2 R+\Theta^2)(\nu^2-3\omega^2)\nn\\
&+&2k(\omega^3\ln R + 4\nu\omega^2\Theta - \nu^2\omega\ln R - 2\nu^3\Theta)\Big)
\Bigg], 
\eea
and 
\bea
{\rm Im}\,\epsilon(\omega,{\bf k}) &=& \frac{m_D^2\omega}{4k^2 G}\Big[2k\Theta-\nu(\ln^2 R+\Theta^2)\Big]
+\frac{m_D^2\xi}{24k^4G}\Bigg[ 6\omega k^2\cos 2\eta (2k\Theta-\nu(\ln^2 R+\Theta^2))-(1+3\cos 2\eta)\nn\\
&\times&\Big(2k^3\nu\ln R+24k^2\omega\nu+3\omega\nu(3\nu^2-\omega^2)(\ln^2 R+\Theta^2)
+6k\omega(-2\omega\nu\ln R + \Theta(\omega^2-5\nu^2))\Big)
\Bigg]
\eea
where
$R = \sqrt{\frac{(\omega + k)^2 + \nu^2}{(\omega - k)^2 + \nu^2}}$, 
$\Theta = \arccos\frac{\omega^2-k^2+\nu^2}{\sqrt{(\omega^2-k^2+\nu^2)^2+4k^2\nu^2}}$ and 
$G = (k-\frac{\nu\Theta}{2})^2+\frac{\nu^2}{4}\ln^2 R$.\\
It is important to note that ${\rm Re}\,\epsilon$ gives the dispersion relation of the 
collective modes and ${\rm Im}\,\epsilon$ gives the damping or attenuation of these modes.

\section{Wake Potential}
In this section we derive the wake potential for collisional
anisotropic($\nu\ne 0, \xi\ne0$) plasma and compare it with 
collisionless anisotropic plasma ($\nu = 0, \xi\ne0$),
collisional isotropic plasma($ \nu\ne 0, \xi = 0$) as well as
collisionless isotropic plasma ($\nu = 0, \xi = 0$).
The wake potential induced by the fast parton can be obtained 
from the Poisson equation as~\cite{ichimaru}:
\be
\Phi^a({\bf k},\omega)=\frac{\rho^a_{\rm ext}({\bf k},\omega)}{k^2\epsilon({\bf k},\omega)}\label{phi}.
\ee
where $\rho^a_{\rm ext}({\bf k},\omega)$ is external color charge density.

Now we study the wake behavior of QGP interacting with 
a charge particle, $Q^a$ moving with  constant velocity 
${\bf v}$. The external color charge density associated 
with the test charge particle can be written as~\cite{ichimaru},
\be
\rho^a_{\rm ext} = 2\pi Q^a\delta(\omega-{\bf k.v}).\label{ext}
\ee 
The wake potential in configuration space due to the motion 
of a charge parton can be calculated with the help of 
Eqs.~(\ref{phi}) and ~(\ref{ext}) and it reads as
\be
\Phi^a({\bf r},t) = 2\pi Q^a\int \frac{d^3k}{(2\pi)^3}\int 
\frac{d\omega}{2\pi}~{\rm exp}^{i({\bf k.r}-\omega t)}
\frac{1}{k^2\epsilon(\omega,{\bf k})}\delta(\omega-{\bf k.v}).
\ee
We evaluate the wake potential for the two special cases, 
(a) along the parallel direction of  motion of 
the parton, i.e., ${\bf r}~||~{\bf v}$ and (b) perpendicular to 
the direction of motion of the parton, i.e., ${\bf r}\perp {\bf v}$.
\subsection{Wake potential along the parallel direction}
The wake potential for the parallel case becomes
\bea
\Phi^a_{||}({\bf r},t) = Q^a m_D\int \frac{d^3k}{(2\pi)^3}
\Big[\cos\Gamma\frac{{\rm Re}\,\epsilon({\bf k},\omega)}{\Delta} 
+ \sin\Gamma\frac{{\rm Im}\,\epsilon({\bf k},\omega)}{\Delta}\Big]\Bigg{\rvert}_{\omega = {\bf k.v}}
\eea
where $\Gamma = m_D k (r-v\,t)\cos\theta$ and 
$\Delta = ({\rm Re}\,\epsilon({\bf k},\omega))^2 + 
({\rm Im}\,\epsilon({\bf k},\omega))^2$. 
Note that $\cos\eta = \cos\lambda\cos\theta + 
\sin\lambda\sin\theta\cos\phi$, where
$\lambda$ is the angle between ${\bf r}$ and $z$ and $\theta$, 
$\phi$ are the polar and azimuthal angles corresponding to
${\bf k}$.

\begin{figure}[t]
\begin{center}
\epsfig{file=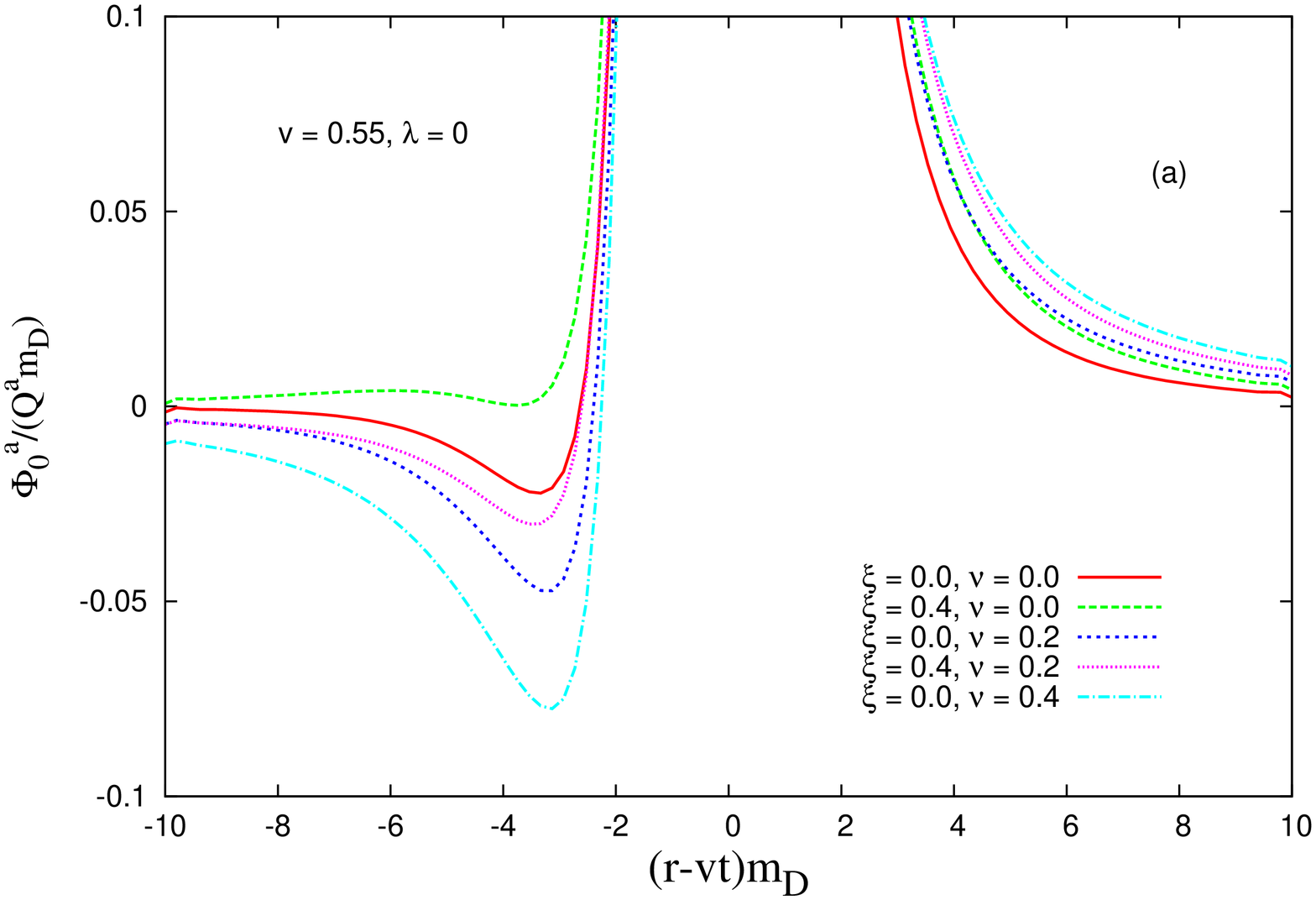,width=7cm,height=7cm,angle=270}~
\epsfig{file=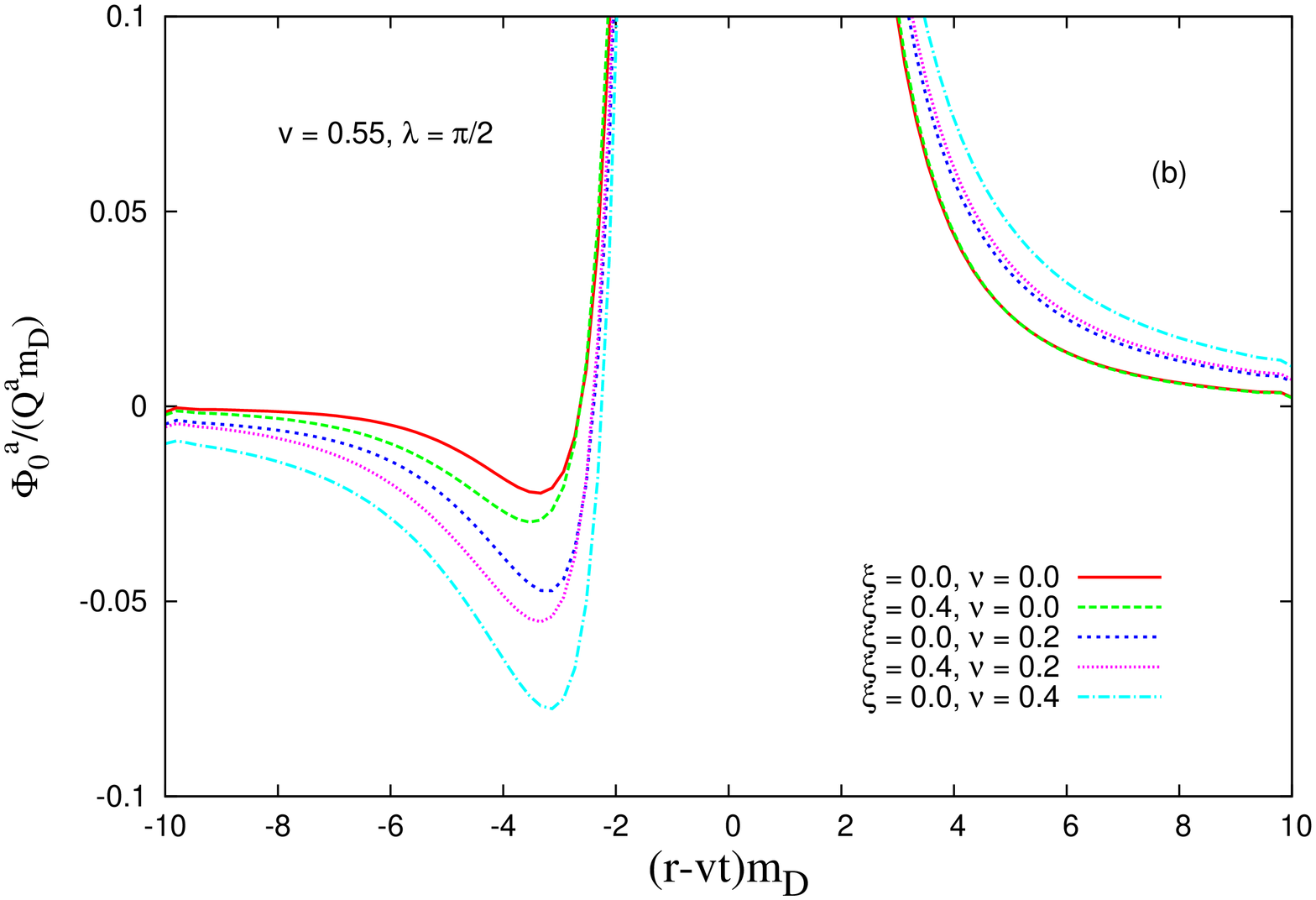,width=7cm,height=7cm,angle=270}
\end{center}
\caption{(Color online) The scaled wake potential along the direction of 
motion of parton for different $\lambda \{= 0,\pi/2\}$
with parton velocity $v = 0.55$
}  
\label{fig1}
\end{figure}

\begin{figure}[t]
\begin{center}
\epsfig{file=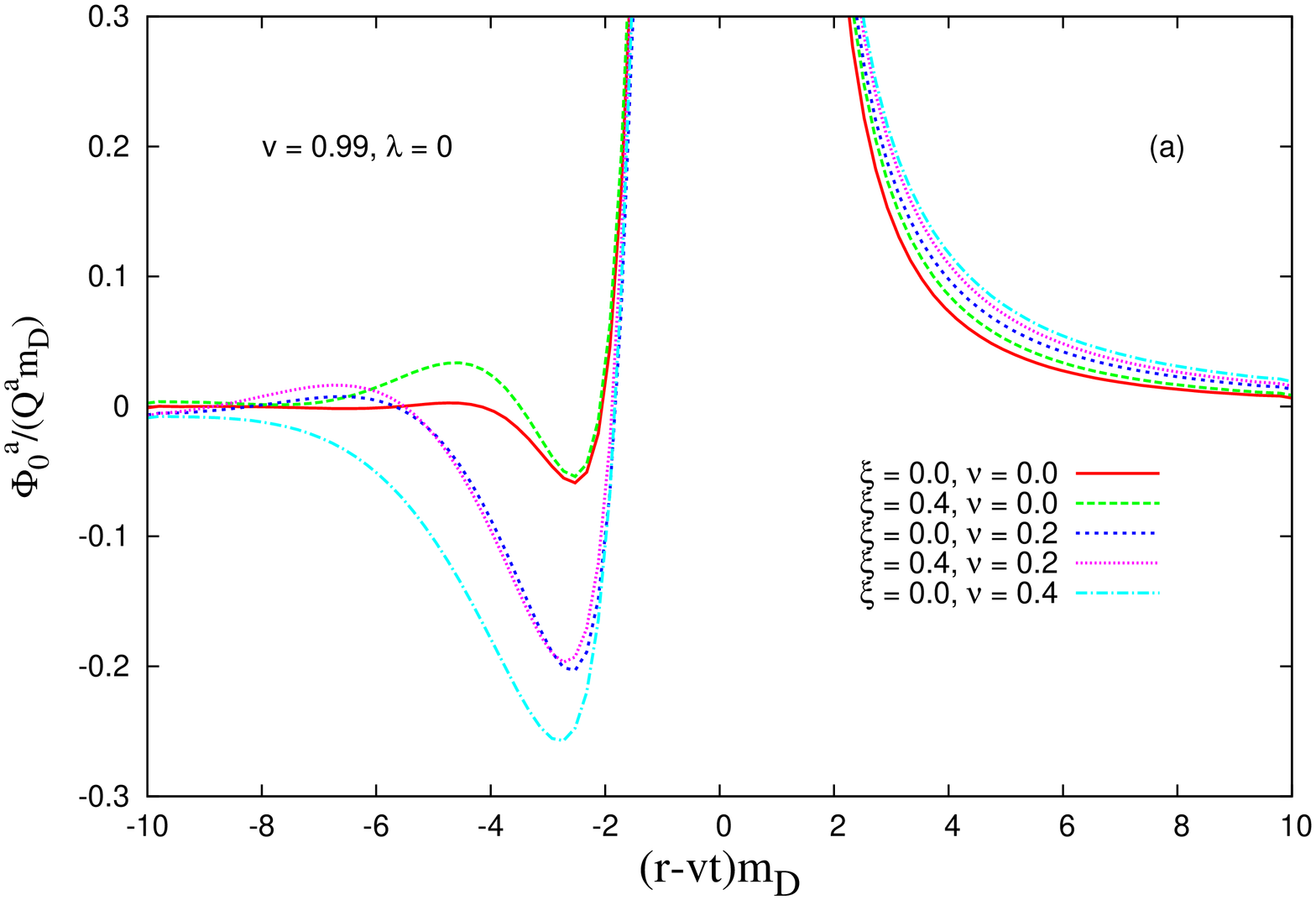,width=7cm,height=7cm,angle=270}~
\epsfig{file=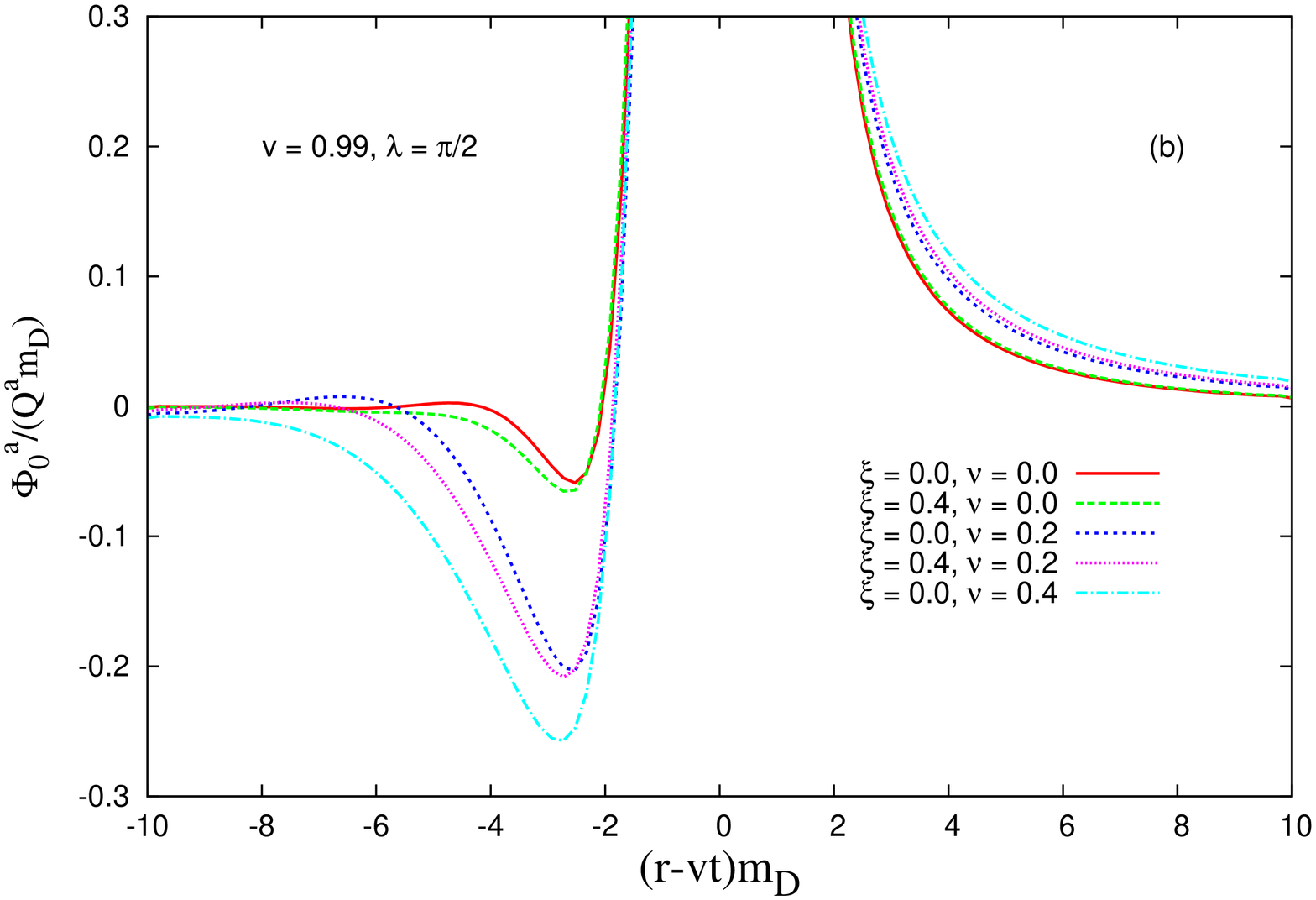,width=7cm,height=7cm,angle=270}
\end{center}
\caption{(Color online) Same as Fig.~\ref{fig1} with $v = 0.99$ 
}  
\label{fig2}
\end{figure}

\begin{figure}[t]
\begin{center}
\epsfig{file=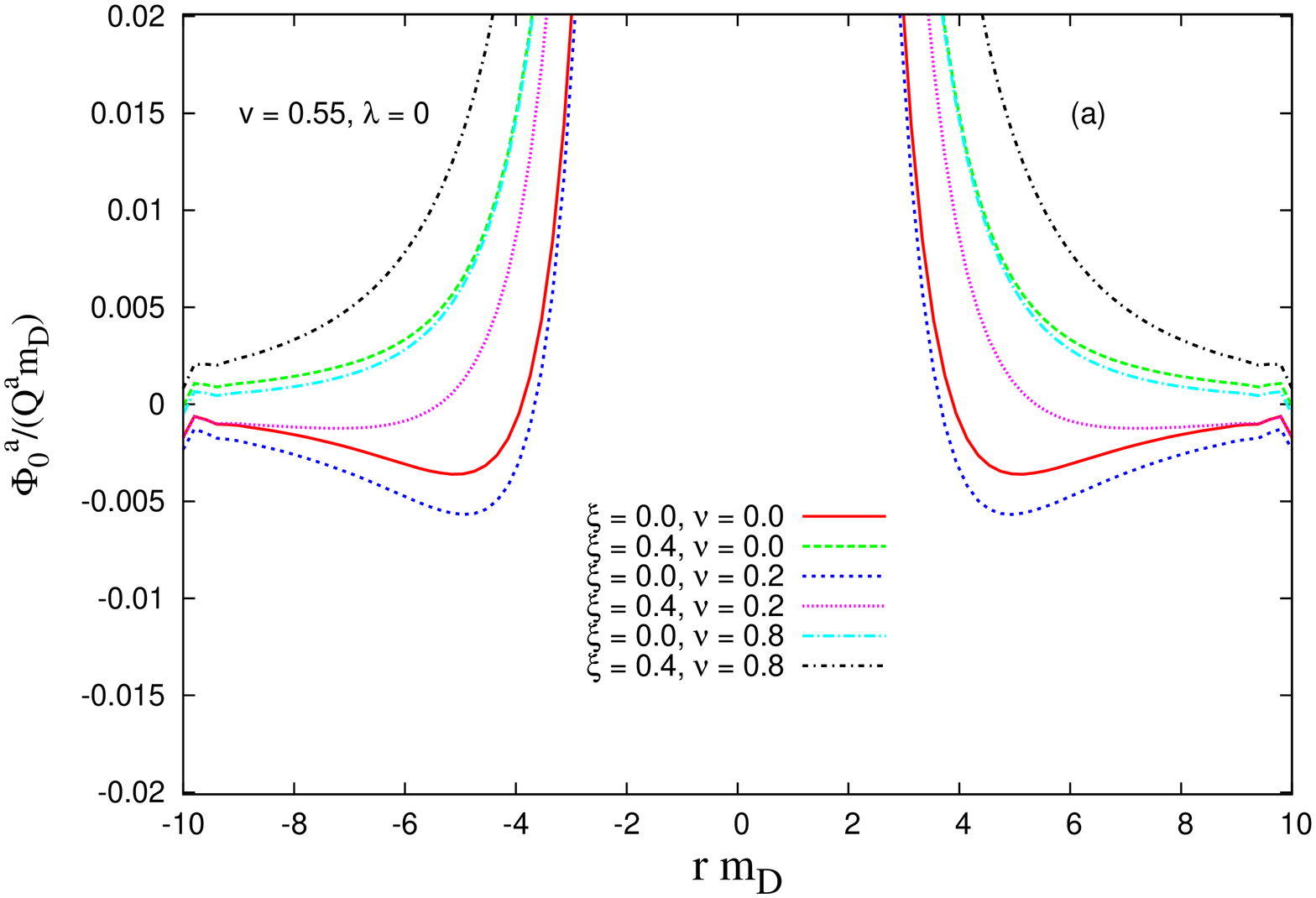,width=8cm,height=8cm,angle=270}~
\epsfig{file=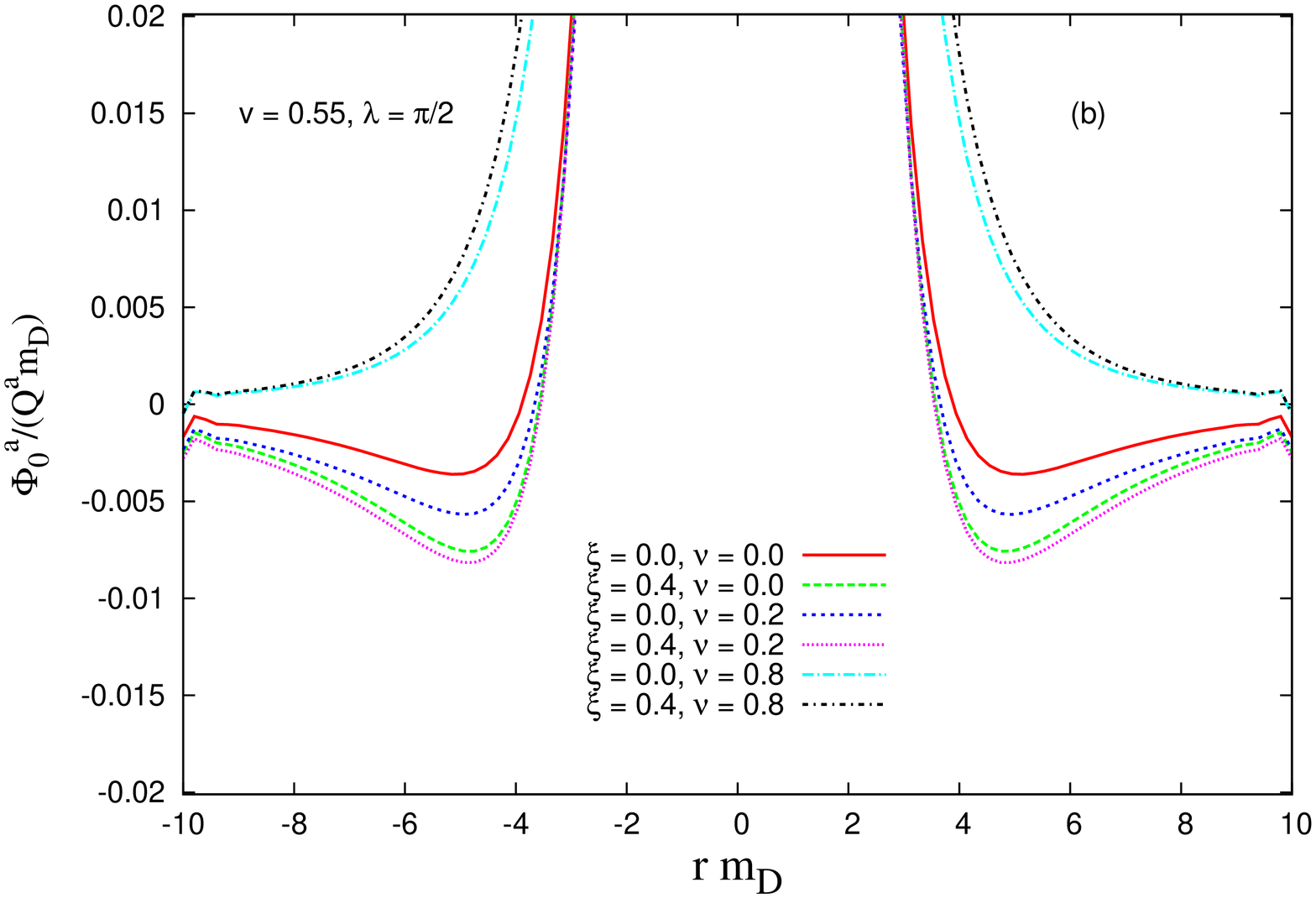,width=8cm,height=8cm,angle=270}
\end{center}
\caption{(Color online) The scaled wake potential along perpendicular direction of 
motion of parton for different $\lambda \{= 0,\pi/2\}$
with parton velocity $v = 0.55$}  
\label{fig3}
\end{figure}

Here we present the numerical evaluation of the wake potential
in parallel direction (${\bf r}||{\bf v}$) for two values of $\lambda$
i.e. $\lambda = 0$ and $\lambda = \pi/2$, where $\lambda = 0 (\pi/2)$
corresponds to ${\bf r}||{\bf {\hat{n}}} ({\bf r}\perp{\bf {\hat{n}}})$.
Numerical results of the wake potential along the parallel 
direction (${\bf r} \parallel {\bf v}$) of motion of the parton 
are shown in Figs~(\ref{fig1})
and~(\ref{fig2}) with parton velocities $v = 0.55$ and $v = 0.99$,
respectively. In these figures, the scaled parameter $\Phi^a_0$ is given 
by $\frac{2\pi^2}{m_D}\Phi^a$. The left (right) panels in Figs.~(\ref{fig1}) 
and~(\ref{fig2}) represent  the case for $\lambda = 0~(\pi/2)$ 
i.e. ${\bf r}\, ||\, {\bf \hat n}$ (${\bf r} \perp {\bf \hat n}$).

In Fig. (1a) we presents the wake potential when $\lambda = 0$, $v = 0.55$ and 
$(r\,-\,v\,t)<\,0$ (backward direction). It is seen that for $\xi = 0$,
$\nu = 0$ (collisionless isotropic plasma), the potential becomes 
Lennard-Jones type. For moderate values of $\xi$ with $\nu = 0$, the potential
still remains as Lennard-Jones type, however the depth of the negative 
minimum decreases. If the value of $\xi$ is increased, the potential 
changes to modified Coulomb-like potential.  This is the unique feature
that we observe in this work. It is also seen that for collisional 
($\nu = 0.2$) isotropic plasma, the depth of the negative minima increases
compared to the isotropic case with $\nu = 0$. We also observe that when 
both the parameters
are nonzero, the potential remains Lennard-Jones type and the depth of the 
negative minima decreases in these cases. We thus concluded that because 
of the variation of the depth of the potential in the cases considered
here there 
will be substantial effect on the observables such as heavy quarkonium 
suppression. Next we consider the behavior of the wake potential when 
$\lambda = 0$, $v = 0.99$ and $(r\,-\,v\,t)<\,0$. This is shown in 
Fig.(2a). For isotropic collisionless plasma($\xi = 0$, $\nu = 0$) 
the potential is of Lennard-Jones type  with oscillatory behavior. 
When $\xi = 0.4$ and $\nu = 0$, the form of the potential does not change.
However, it becomes more oscillatory compared to the previous case for the 
reason explained in ~\cite{prd86}. It is important to note that the depth of 
the negative minima remains unchanged in these cases. For collisional isotropic
plasma($\nu\,\ne\,0$, $\xi =\,0$), the depth of the potential increases as
compared to the case of collisionless isotropic plasma. The form of the 
potential 
remains unchanged with oscillatory behavior. If we increase the collision 
frequency, the depth of the potential becomes highest and most importantly
the oscillatory behavior is smeared out. The combination 
($\xi = 0.4$, $\nu = 0.2$)
leads to more oscillatory potential as compared to the case when 
$\xi = 0$ and $\nu = 0.2$. The overall observations in this case 
are as follows: The potential remains Lennard-Jones type in all 
the cases considered here. The depth of the negative minima varies 
depending upon the combinations of $\nu$ and $\xi$. For large 
value of $\nu$, if we increase $\xi$ the oscillatory nature of 
the potential is washed away. 

We now turn our attention to the case when $\lambda = \pi/2$
i.e. ${\bf r}\perp {\bf\hat n}$ and $v = 0.55$. The result is
displayed in Fig. (1b), various combinations
of $\xi$ and $\nu$ have been considered. In all the cases the 
potential is Lennard-Jones type and the depth of the negative
minima is larger than the isotropic case if one of the parameters
is nonzero or both are nonzero. Thus in this scenario the 
quarkonium state will be strongly bound and therefore, the 
dissociation temperature required to break the state will be higher
as compared to the isotropic case. Moreover, it will certainly 
affect the conical flow and other observables. The wake potential
corresponding to the previous case with  larger velocity is shown
in the Fig. (2b). As before, for collisionless isotropic plasma,
the potential is of Lennard-Jones type with oscillatory behavior.
With nonzero $\xi$, the oscillatory behavior vanishes 
contrary to the case when ${\bf r}\parallel {\bf\hat n}$ for a 
collisionless isotropic plasma. For collisional($\nu = 0.2$) isotropic
plasma, the oscillatory behavior is observed again. If we increase 
$\nu$, the oscillatory behavior again vanishes. The vanishing 
and reappearing of the oscillatory behavior for certain combinations
of $\xi$ and $\nu$ is because of the fact that there is subtle competition
between collisional isotropic plasma and collisionless anisotropic
plasma. Finally we note that in the 
forward direction in all  the above cases, the wake potential is a
modified Coulomb-like potential.

We have also examined the case for large $\xi$. For $v = 0.55$ and 
$\lambda =0 $, the potential becomes coulomb-like. It is also seen
that increasing the value of $\xi$ the effect of collision becomes
unimportant. However, for faster speed the potential remains Lennard-Jones
type with no oscillatory behavior in case of collisional plasma.
For $\lambda = \pi/2$, the potentials remain of Lennard-Jones
type for both values of $v$ considered here.

\subsection{Wake potential along the perpendicular direction}
Next we consider the wake potential for the perpendicular case which reads as:
\bea
\Phi^a_{\perp}({\bf r},t) = Q^a m_D\int \frac{d^3k}{(2\pi)^3}
\Big[\cos\Gamma^{\prime}\frac{{\rm Re}\,\epsilon({\bf k},\omega)}{\Delta}
+ \sin\Gamma^{\prime}\frac{{\rm Im}\,\epsilon({\bf k},\omega)}{\Delta}\Big]\Bigg{\rvert}_{\omega = {\bf k.v}}
\label{pot_perpedicular}\eea
where $\Gamma^{\prime} = m_D k(r\cos\theta-v\,t\sin\theta\cos\phi)$.

\begin{figure}[t]
\begin{center}
\epsfig{file=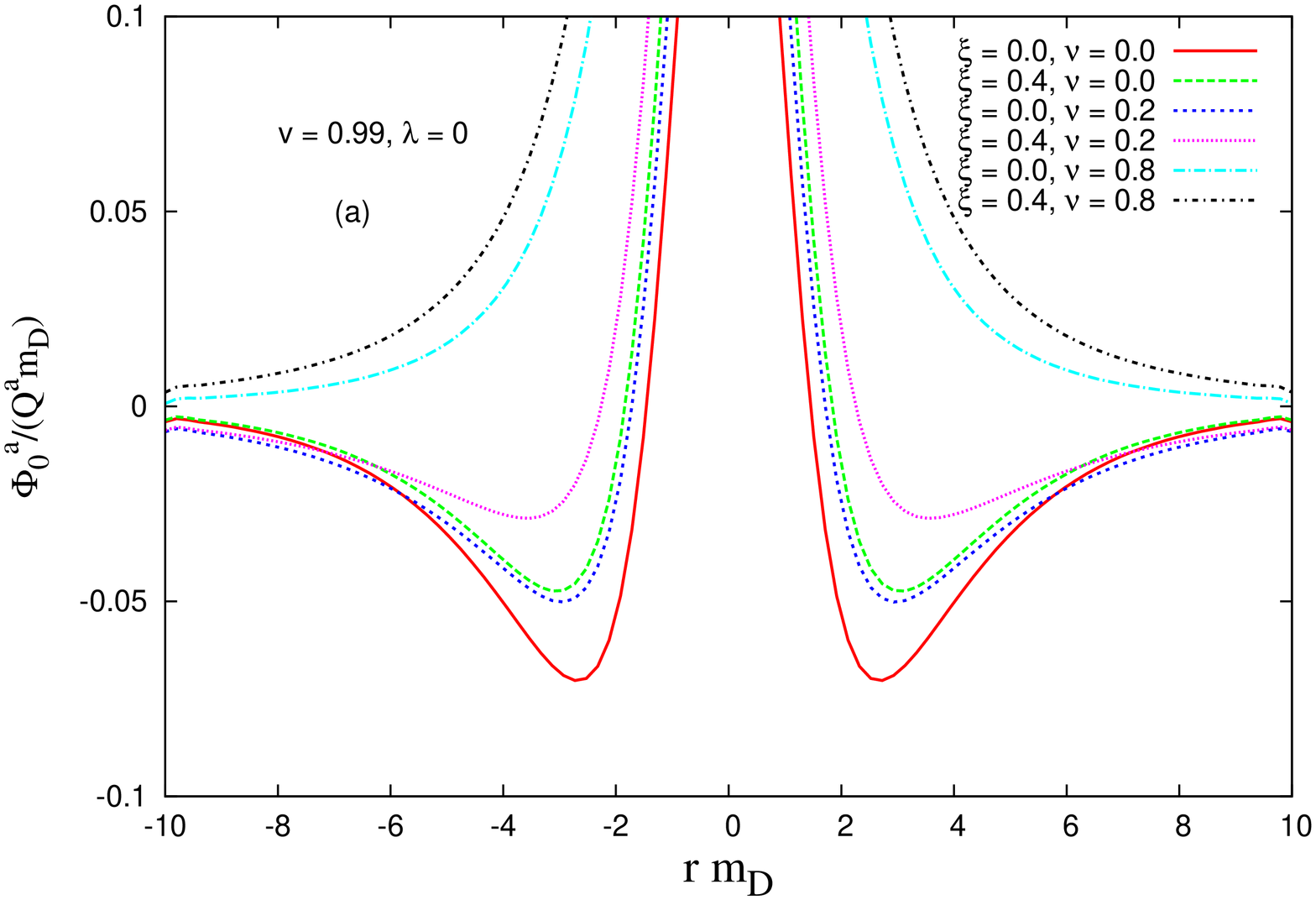,width=8cm,height=8cm,angle=270}~
\epsfig{file=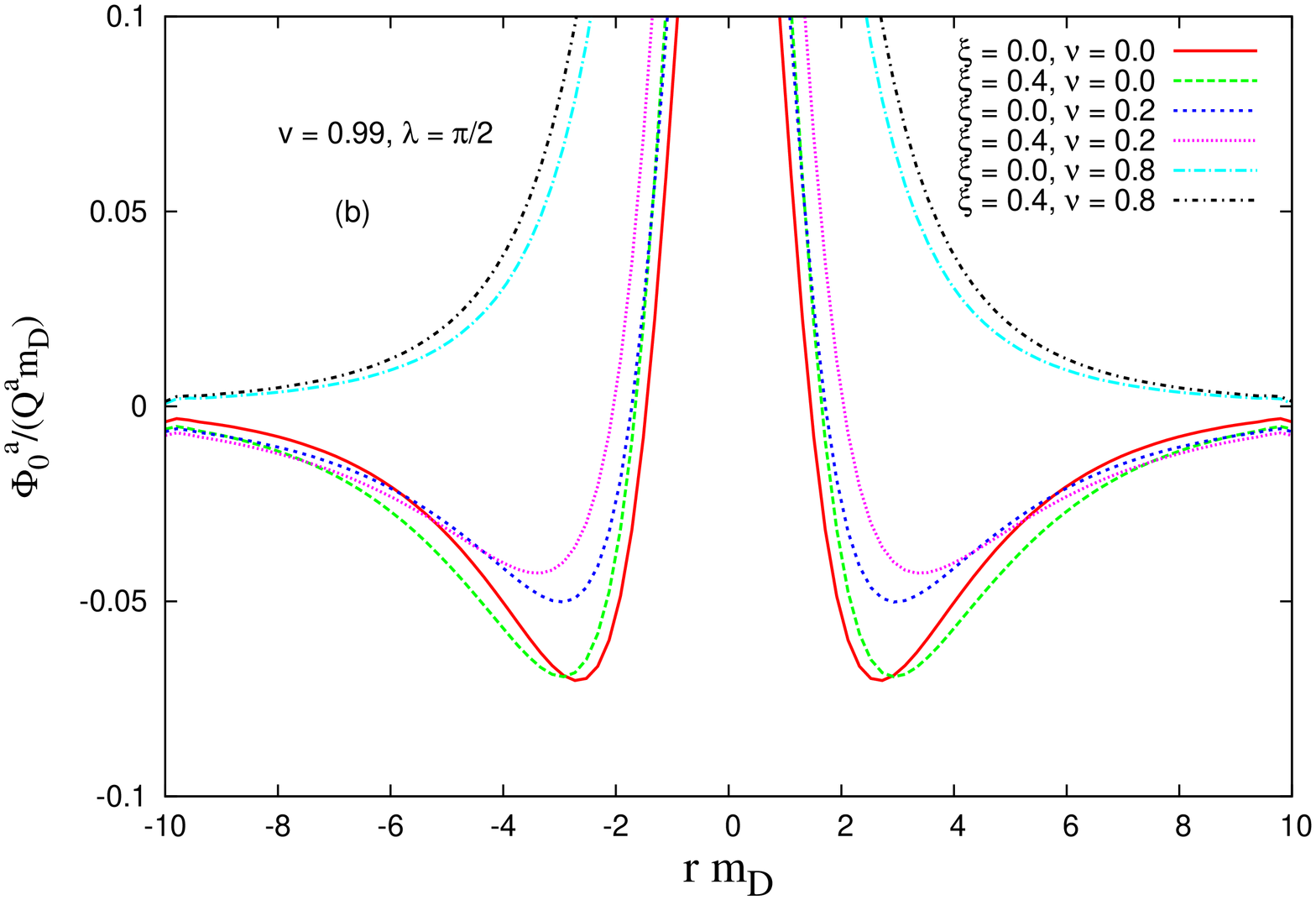,width=8cm,height=8cm,angle=270}
\end{center}
\caption{(Color online) Same as Fig.~\ref{fig3} with $v = 0.99$ }  
\label{fig4}
\end{figure}

We numerically evaluate Eq.(\ref{pot_perpedicular}) and the corresponding results (scaled
wake potential) of the wake 
potential along the perpendicular direction of motion of the parton 
(${\bf r}\perp {\bf v}$) are shown in
Figs.~(\ref{fig3}) and (\ref{fig4}) for 
$v = 0.55$ and 
$v = 0.99$ respectively.
We first discuss the case when $\lambda = 0({\bf r}\parallel{\bf \hat n})$, 
and  $v = 0.55$. The corresponding results are shown in the left panel 
of Fig (3). It is seen that the wake potential is forward-backward symmetric.
Let us first concentrate on  the case of collisional isotropic plasma. For $\nu = 0.2$
the potential is of Lennard-Jones type as in the collisional isotropic plasma.
However, the depth of the negative minima increases. If we increase $\nu (= 0.8)$, 
the potential turns into a modified Coulomb -like potential. However, for $\xi\,\ne\,0$ 
(anisotropic plasma), the wake potential remains modified Coulomb-like potential
irrespective of the value of the collisional frequency($\nu$). On the 
other hand, for $\lambda = \pi/2$(${\bf r}\perp{\bf \hat n}$), the 
wake potential remains Lennard-Jones type up to $\nu = 0.2$ and $\xi = 0.4$
with the depth of the negative minima increasing with $\xi$ and $\nu$(see 
right panel of Fig. (3). For large values of $\nu(0.8)$, the wake 
potential turns into Coulomb-like potential.

For the large velocity i.e. for $v= 0.99$, we find no oscillatory behavior
for collisionless isotropic plasma for $\lambda = 0({\bf r}\parallel{\bf \hat n})$.
(see left panel of Fig. (4)). Thus the wake potential created by the 
parton moving in transverse plane has different behavior than when it 
moves in the parallel direction. This should be reflected in the 
observables which have origin in the moving parton(such as di-hadron 
correlations). The results for $\lambda = \pi/2({\bf r}\perp{\bf \hat n})$
and $v = 0.99$ are shown in the right panel of Fig. (4). The potential
remain of Lennard-Jones type for various combinations of $\xi$ and $\nu$ as 
long as collisional frequency is small. When $\nu = 0.8$, the potential
turns into modified Coulomb-like potential. It is also seen that the depth of the 
negative minima decreases irrespective of the value of $\xi$.

\section{Summary}
In this work we have investigated the behavior of the wake
potential induced by a fast parton propagating through the 
collisional quark-gluon plasma which is 
anisotropic in momentum space. To simplify the analysis, we have
calculated the dielectric response function in small $\xi$
limit including the effect of BGK collisional kernel. We have 
presented the wake potentials for both  
parallel (${\bf r}~||~{\bf v}$) and perpendicular (${\bf r}\perp {\bf v}$) 
directions of motion of the fast parton
with two different parton velocities. We see that when the parton
moves in the parallel(${\bf r}~\parallel~{\bf v}$) direction with $v = 0.55$, the potentials
become of Lennard-Jones type in the backward region and 
Coulomb-like in the forward region for $\lambda = 0 
({\bf r}\parallel {\bf\hat n})$ and $\lambda = \pi/2
({\bf r}\perp {\bf\hat n})$. However, the depth of the negative minimum
varies with $\xi$ and $\nu$. For $v = 0.99$ the potential shows less
oscillatory than collisionless isotropic and anisotropic QGP. 
We have also evaluated the
wake potentials when ${\bf r}\perp {\bf v}$. In this case when
$v = 0.55$, ${\bf r}\parallel {\bf\hat n}\,(\lambda = 0)$ and
${\bf r}\perp {\bf\hat n}\,(\lambda = \pi/2)$ it is observed that the
potentials remain of Lennard-Jones type for
$\nu$ up to 0.2 and for various values of $\xi$ (within the small $\xi$ limit).
However, if we increase $\nu$ the potential becomes Coulomb-like 
irrespective of the values of $\xi$. 
For $v = 0.99$, ${\bf r}\parallel {\bf\hat n}\,(\lambda = 0)$ and 
${\bf r}\perp {\bf\hat n}\,(\lambda = \pi/2)$ more or less similar
behavior of the wake potential has been observed as in the
case for $v = 0.55$ . However, with any combinations of
$\xi$ and $\nu$  we do not find any oscillatory behavior in
the wake potential. 
Observations of the oscillatory and non-oscillatory
behaviors of the wake potential for certain combinations of
$\xi$ and $\nu$ imply that, as long as $xi$ remains below a certain
critical value, the effect of collision becomes dominant. But
when $xi$ is greater than this critical value, the anisotropy
effect takes over, irrespective of the values of $\nu$, and the
wake structure becomes similar to the case of collisionless
AQGP. We also note that the anisotropic effect on the wake
potential is speed dependent, and it has rich structure com-
pared to the case of collisionless isotropic plasma. We
observe that for $v = 0.99$ the oscillatory effect is more
pronounced for $\xi\neq 0$ and $\nu = 0$. However, it decreases
with the inclusion of collision. But for lower speeds the
effect of anisotropy on the wake potential is minimal if the
collision is included. We shall end by mentioning some of
the phenomenological implications of the present work. In
the present calculation, we see that the wake structure for
collisional AQGP is less pronounced compared to the case
of purely isotropic plasma, as well as collisionless AQGP.
This will surely influence the conical flow and wave excitation 
in the plasma. Second, as the depth of the potential
changes depending upon the values of $\xi$ and $\nu$, it can change
the repulsive and attractive parts of the interaction potential
of a fast parton in the plasma. In particular, in our work, we
have shown that the depth of the negative minima becomes
deeper for collisional anisotropic QGP as well as collisional
isotropic QGP when ${\bf r}\parallel {\bf\hat n}$. This will lead to a stronger
binding of diquarks propagating through the plasma requiring 
larger values of the dissociation temperature to separate
the heavy quark bound state in such a scenario. Thus, the
nuclear modification factors of charm and bottom mesons
should be reevaluated, applying the concept of the present
work. Moreover, it will also be interesting to explore
whether such changes in the wake potential can influence
the dihadron correlations.

\appendix
\section{Derivation of Eq.~(\ref{di-self})}
In this section we derive Eq.~(\ref{di-self}). For this, we note that the 
spatial component of induced current density is given by
(see Eq.~(\ref{currentdensity})),
\bea
J^i_{ind}(K) &=& g^2\int_{{\bf p}} V^i\partial^l_{(p)}f({\bf p}){\cal M}_{jl}(K,V)D^{-1}
(K,{\bf v},\nu)A^j(K) + 2N_cg\nu{\cal S}^g(K,\nu) \nn\\
&+& g^2(i\nu)\int\frac{d\Omega}{4\pi}
V^iD^{-1}(K,{\bf v},\nu)\int_{{\bf p}^{\prime}}\partial^l_{p^{\prime}}f({\bf p}^{\prime})
{\cal M}_{jl}(K,V^{\prime})D^{-1}(K,{\bf v}^{\prime},\nu){\cal W}^{-1}(K,\nu)A^j(K)\nn\\
&+& 2N_cg^2(i\nu^2)\int\frac{d\Omega}{4\pi}V^iD^{-1}(K,{\bf v},\nu){\cal S}^g(K,\nu)
{\cal W}(K,\nu).~\label{A1}
\eea
The spatial component of the polarization tensor can be written from 
Eq.~(\ref{polarization})
as:
\bea
\Pi^{ij}(K) &=& g^2\int_{{\bf p}} V^i\partial^l_{(p)}f({\bf p}){\cal M}_{jl}(K,V)D^{-1}
(K,{\bf v},\nu)\nn\\
&+&  g^2(i\nu)\int\frac{d\Omega}{4\pi}
V^iD^{-1}(K,{\bf v},\nu)\int_{{\bf p}^{\prime}}\partial^l_{p^{\prime}}f({\bf p}^{\prime})
{\cal M}_{jl}(K,V^{\prime})D^{-1}(K,{\bf v}^{\prime},\nu){\cal W}^{-1}(K,\nu)
\eea
We also note that the thermal conductivity and 
the dielectric tensor is related by~\cite{canj82}
\be
\epsilon^{ij}(K) = \delta^{ij} + \frac{i}{\omega}\sigma^{ij}(K)
\ee
where $\sigma^{ij}(K) = \frac{\delta J^i_{ind}(K)}{\delta E_j(K)}$
Using Eq.~\ref{A1} we derive
\bea
\sigma^{ij}(K) &=& \frac{ig^2}{\omega}\Bigg[\int_{{\bf p}} V^i\partial^l_{(p)}
f({\bf p}){\cal M}_{jl}(K,V)D^{-1}(K,{\bf v},\nu)\nn\\
&+&i\nu\int\frac{d\Omega}{4\pi}V^iD^{-1}(K,{\bf v},\nu)\int_{{\bf p}^{\prime}}
\partial^l_{p^{\prime}}f({\bf p}^{\prime})
{\cal M}_{jl}(K,V^{\prime})D^{-1}(K,{\bf v}^{\prime},\nu){\cal W}^{-1}(K,\nu)\Bigg]
\eea
Here we have use the following relations in temporal axial gauge: 
\be
E_i = F_{0i} = \partial_0A_i-\partial_iA_0=-i\omega A_i.
\ee
Using the above relations it is straightforward to show that 
\be
\epsilon^{ij}(K) = \delta^{ij}-\frac{\Pi^{ij}(K)}{\omega^2}.
\ee
The dispersion relation of the collective modes in temporal axial gauge can 
be written as~\cite{prd68}:
\be
{\rm det}\big[(k^2-\omega^2)\delta^{ij}-k^ik^j+\Pi^{ij}(K)\big] = 0
\ee
or equivalently~\cite{prd62}:
\be
{\rm det}\big[k^2\delta^{ij}-k^ik^j-\omega^2\epsilon^{ij}(K)\big] = 0
\ee
Thus, the modes can be obtained either by evaluating $\Pi^{ij}(K)$ or $\epsilon^{ij}(K)$.

\end{document}